\documentclass[12pt,a4paper,notitlepage]{report}

\begin{document}
\centerline{Detection of Neutron Sources in Cargo Containers}

\medskip
\centerline{J. I. Katz}
\centerline{Department of Physics and McDonnell Center for the Space
Sciences}
\centerline{Washington University, St. Louis, Mo. 63130}
\centerline{\today}
\begin{abstract}
We investigate the problem of detecting the presence of
clandestine neutron sources, such as would be produced by nuclear weapons
containing plutonium, within cargo containers.  Small, simple and economical
semiconductor photodiode detectors affixed to the outsides of containers are
capable of producing statistically robust detections of unshielded sources
when their output is integrated over the durations of ocean voyages.  It is
possible to shield such sources with thick layers of neutron-absorbing
material, and to minimize the effects of such absorbers on ambient or
artificial external neutron fluxes by surrounding them with
neutron-reflective material.
\end{abstract}
\bigskip
Terrorist nuclear weapons or special nuclear material may be shipped in
cargo containers, and their detection is a matter of serious concern.  If
the special nuclear material is plutonium then its $^{240}$Pu content is a
significant source of spontaneous fission neutrons, depending on the
quantity of plutonium and its isotopic composition.  These spontaneous
fission neutrons may be detectable.

In an earlier report$^1$ we suggested applying
$^{10}$B-coated photodiode neutron detectors to the outsides of shipping
containers to detect any spontaneous fission neutrons emitted inside.
Simple analytic estimates showed that, integrated over a ten day voyage, a
1 cm$^2$ detector would detect with high statistical significance an
unshielded spontaneous fission neutron source in the presence of the
background of cosmic ray-induced spallation neutrons.  The purpose of this
note is to report the results and implications of quantitative Monte-Carlo
neutron transport calculations of this problem, including the effects of
shielding and the contents of surrounding containers.  These calculations
used the MCNPX code$^{2-4}$, which is a standard tool used to calculate
neutron transport.

In all of these calculations the assumed source was a sphere of 5 kg of
$\delta$-plutonium (radius 4.22 cm).  A source strength of $4.5 \times 10^5$
n/s was chosen (this is appropriate to 10 percent $^{240}$Pu, a compromise
between nominal ``weapons grade'' and ``reactor grade'' compositions).  The
results were normalized to the source rate of spontaneous fission neutrons.
The plutonium was surrounded by a 50 cm thick spherical shell of nominal
``explosive'' (composition C$_7$N$_3$O$_6$H$_5$ and density 1.62 gm$^{-3}$;
TNT).  This is not meant to be a realistic bomb; rather, it was deliberately
chosen as an amateur interpretation of Fat Man (the subject of the Trinity
test and dropped on Nagasaki).

The thick ``explosive'' layer is a fairly effective moderator and reflector
and a significant absorber.  The assembly is close to thermal neutron
criticality (the multiplication factor is 6.0), and for each source neutron
there are 20 neutron crossings of the boundary between plutonium and
``explosive'', while only 0.24 neutrons cross the outer surface of the
``explosive''.

The source was at the center of a 40$^\prime$ shipping container.  The
detector was applied to the surface of the container (the container
walls were ignored) at a point half-way from bottom to top but 3.29 m
displaced from the center of the side of the container along its length, so
that the center of the detector was 3.5 m from the center of the source.
This location was chosen because three detectors may be applied to the
container so that no point within its volume is more than 3.5 m from any
detector.  Three detectors is a reasonable compromise between the
requirements to minimize the number of detectors and to bring every point in
the container as close as possible to a detector.  In order to obtain
accurate statistics with a feasible number of Monte Carlo particles a
disc detector of 1 m radius was used in the calculations, and the inferred
count rate scaled to the more practical 1 cm$^2$ detector.

The detector itself consisted of a 2 $\mu$ layer of boron enriched to 80
percent $^{10}$B.  In all cases the detector was assumed to have an area of
1 cm$^2$.  The silicon or gallium arsenide photodiode, perhaps 10 $\mu$
thick, is not calculated explicitly, but a (conservative) 0.25 efficiency
for detection of (n,$\alpha$) reactions in the boron is assumed, as
discussed by Grober and Katz$^1$.  This layer is sandwiched between two 1.75
cm thick layers of paraffin or polyethylene (CH$_2$ of density 0.92
g/cm$^3$) whose purpose is to thermalize neutrons emitted by a bare source.
When the source is moderated the CH$_2$ acts as a neutron reflector (each
slab has a scattering optical depth, at normal incidence, to thermal
neutrons of 2.8).  The thickness of these layers is constrained by the
corrugations of shipping container walls, which are about 3.75 cm deep.
If the slabs of CH$_2$ were not included the detector would be insensitive
to an {\it unmoderated} neutron source.  A more sophisticated system would
include detectors with and without sandwiching slabs of CH$_2$.

This baseline calculation produces a reaction rate of $1.5 \times 10^{-5}$
cm$^{-3}$ (of the boron film) per source neutron.  In $10^6$ seconds of
exposure (typical of a transoceanic voyage) to a source of
$4.5 \times 10^5$ neutrons s$^{-1}$ a 1 cm$^2$ detector records 340 events,
given the assumed efficiency.  This may be compared to the rough
estimate$^1$ of 600 detected events.  The results cannot be 
compared closely because the estimate assumed a bare unmoderated source
for the baseline problem.

A number of variations on the baseline calculation were performed.  For
example, surrounding the container containing the neutron source with 26
innocent containers (a $3 \times 3 \times 3$ array of containers, with the
neutron source in the central container), with each innocent container
filled with a nominal homogeneous cargo of 0.3 g/cm$^3$ (the known mean
density of container loading), taken to be of composition FeH (3.7\%
hydrogen by mass; only the hydrogen is significant for neutronics so the
composition of the remainder of the mass is irrelevant), increases the
number of (n,$\alpha$) reactions detected to 1600.  This is a consequence of
neutron reflection by the hydrogen in the innocent containers.  It is about
ten times less than the rough estimate$^1$, probably because the latter
ignored the reflective effect of the CH$_2$ slabs around the detector.

The assumption of homogeneous cargo (other than the assumed threat source)
is of uncertain validity and may introduce significant error.  Neutron
transport is very different in a heterogeneous medium, with free passages
between regions of comparatively high hydrogen density, than in a homogeneous
medium.  Some cargoes (for example, a container packed full of clothing)
are reasonably homogeneous, while others (machinery, dense objects with
hydrogenous packing, drums of chemicals) may either contain no hydrogen or
have it concentrated into isolated regions.  Unfortunately, it is probably
not possible to resolve this uncertainty computationally because the
hydrogen distribution in real cargo is difficult to characterize.  
Straightforward experiments in which real neutron sources (D-D accelerators,
($\alpha$,n) or $^{252}$Californium, not plutonium!) are placed in
instrumented containers among innocent cargo are probably necessary to
resolve these uncertainties.

These count rates should be compared to those produced by cosmic ray
spallation neutrons.  The chief source of these neutrons is spallation in
the cargo (both the threat source and surrounding innocent cargo), rather
than in the air, because cargo densities (0.3 g/cm$^3$, of which very little
is hydrogen) are hundreds of times greater than the density of air.  The
production of neutrons by cosmic ray interactions in surrounding cargo and
the ship itself is known as the ``ship effect''.  Solar neutrons are
negligible compared to those made by spallation.

Again, the FeH composition was assumed; the spallation neutron production
rate is nearly independent of the composition of the non-hydrogenous portion
of the material.  Therefore, MCNPX calculations were done in which
containers containing innocent cargo were irradiated by cosmic rays, and
neutron production, transport and reaction were calculated.  The sea-level
cosmic ray spectrum from 100 MeV to 100 GeV was taken from Pal$^5$, and
vertical downward incidence distributed uniformly across the top of the
container at a flux of $2.7 \times 10^{-4}$ cm$^{-2}$ s$^{-1}$ was
assumed (this makes it a reasonable approximation to ignore cosmic rays
entering the container from the side).  The cosmic rays were assumed to be
all protons because the more abundant muons rarely produce neutrons.

The result for irradiation of a single container filled with the same
model (FeH at 0.3 g cm$^{-3}$) of innocent cargo, integrated over $10^6$ s,
with the same detector as before, was 0.6 detected reaction.  When an array
of 27 similar containers was taken this increased to 1.7 detected reactions,
as a result of reflection of neutrons by the surrounding containers.  These
results are very small compared to the predicted signals of the assumed
threats discussed above, and indicate that the natural neutron background
is not a significant obstacle to detection of spontaneous fission neutrons
from the assumed threat.  These backgrounds are about two orders of magnitude
less than the rough estimates of Grober and Katz$^1$, in part because
the estimates did not allow for thermal neutron reflection by the CH$_2$ 
slabs surrounding the detector, and in part because of differing assumptions 
regarding the cosmic ray flux (equivalently, geomagnetic latitude).

Unfortunately, anyone capable of building even an improvised plutonium-based
nuclear device is probably aware of the desirability of shielding the
spontaneous fission neutrons and is capable of doing so.  An effective shield
is a 50 cm layer of borated polyethylene or paraffin.  A series of
calculations was run with such a shield, in the form of a spherical shell of
inner radius 54.22 cm and outer radius 104.22 cm containing 5\% natural boron
by mass.  The result was to reduce the detected counts in the baseline
calculation to 0.08, an effective shielding factor of about 4000.  This is
clearly undetectable.  If the source is surrounded by 26 containers filled
with innocent cargo 0.7 counts are detected, the effective shielding factor
is 2000, and the source would still be undetectable.

Thermal neutron absorbers produce a ``hole'' in the surrounding distribution
of thermal neutrons.  Would it be possible to detect the presence of such
an absorber (comparatively uncommon in innocent cargo) by measuring the
density of thermal neutrons produced by cosmic rays?  Two calculations were
done to test this.  In the first the absorber (with only a void inside)
was placed in a container, and the container outside the absorber was filled
with the usual assumed FeH innocent cargo.  In the second calculation this
container was surrounded by 26 containers each filled with innocent cargo.

The results agreed, to within a few percent, with the results for containers
containing innocent cargo but no neutron absorbers.  The reason for this is
that once neutrons have thermalized (the detector is not sensitive to
unthermalized neutrons) their scattering mean free path in the cargo is
about 7.5 cm.  The detector, about 2.5 m (33 mean free paths) from the
absorber, senses a neutron field essentially unaffected by the presence of
the absorber.  Of course, someone deliberately hiding the presence of
absorber would surround it with material of even greater hydrogen density
(for example, CH$_2$, in which the mean free path is about 0.6 cm).  Hence
thermal neutron detection is not likely to be a feasible method of 
detecting thermal neutron absorbers if even the most minimal measures are
taken to conceal them within a neutron-scattering medium.

Similar conclusions apply to interrogation of containers with fast neutrons.
Although they (like the spontaneous fission neutrons calculated here) are
much more penetrating than thermal neutrons, it is still possible, within
the volume of a cargo container, to moderate and absorb them.  Even if
fission were induced in fissionable material (by a penetrating interrogation
beam, or the natural muon background), that material could be surrounded by
enough moderator and absorber to prevent detection of the fission neutrons,
and by enough neutron scatterer to prevent detection of the neutron
absorber.

We conclude that boron neutron detectors can be an effective means of 
detecting the presence of unshielded neutron sources, such as kilogram
quantities of plutonium, within cargo containers (uranium, of any isotopic
composition, is a negligible source of neutrons).  A knowledgeable adversary
may shield such sources, and the presence of such shields may be concealed
from detection by their effects on the background thermal neutron density if
they are surrounded with neutron-reflective material.

\bigskip
\centerline{References}

\medskip
\begin{enumerate}
\item Grober, R. D. and Katz, J. I., MITRE/JASON JSR-02-340 Radiological
Weapons Appendix VI (2002).
\item Hughes, H. G., {\it et al.}, LA-CP-02-408, Los Alamos National
Laboratory, Los Alamos, N. Mex. (2002a).
\item Hughes, H. G., {\it et al.}, LA-UR-02-2607, Los Alamos National
Laboratory, Los Alamos, N. Mex. (2002b).
\item Hendricks, J. S., {\it et al.}, LA-UR-04-0570, Los Alamos National
Laboratory, Los Alamos, N. Mex. (2004).  
\item Pal, Y., Handbook of Physics 2nd ed., E. U. Condon and H. Odishaw, eds.,
McGraw-Hill (1967) Fig. 11.22.
\end{enumerate}
\bigskip
\newpage
\centerline{Table of Problems}

\medskip
\begin{tabular}{rlc}
No. & Features & Counts in $10^6$ s \\
1 & Baseline (4.22 cm Pu, 50 cm HE) & 340 \\
2 & Baseline + 26 innocent containers & 1600 \\
3 & Innocent container with cosmic rays & 0.6 \\
4 & 27 innocent containers with cosmic rays & 1.7 \\
5 & Baseline with 50 cm borated CH$_2$ shield & 0.08 \\
6 & Baseline + shield + 26 innocent containers & 0.7 \\
7 & No. 3 + shield & 0.6 \\
8 & No. 4 + shield & 1.8 \\
\end{tabular}
\end{document}